\documentclass[useAMS,usenatbib]{aastex}
\usepackage{spr-astr-addons}
\usepackage{url}
\usepackage[T1]{fontenc}
\usepackage{subfloat}
\usepackage{graphicx}
\RequirePackage{color}

\begin{document}

\title{Investigating the coronal structure by studying time lags in the Atoll source 4U 1705-44 using AstroSat}
\shorttitle{Understanding coronal structure using time lags}
\shortauthors{Malu et al.}

\author{S. Malu \textsuperscript{1,*}, S. Harikrishna \textsuperscript{1}, K. Sriram \textsuperscript{1}, Vivek. K. Agrawal \textsuperscript{2}}
\altaffiltext{*}{malu.sudhaj@gmail.com}
\affil{\textsuperscript{1}Department of Astronomy, Osmania University, Hyderabad, 500007, India.\\}
\affil{\textsuperscript{2}Space Astronomy Group, ISITE Campus,U R Rao Satellite Center, Bangalore, 560037, India.\\}

\begin{abstract}
We performed a detailed timing study of the Atoll source 4U 1705-44 in order to 
understand the accretion disk geometry. Cross correlation function (CCF) studies 
were performed between soft (3-5 keV) and hard energy (15-30 keV) bands using the 
AstroSat LAXPC data. We detected hard as well as soft lags of the order of few ten 
to hundred seconds. A dynamical CCF study was performed in the same energy bands for one of the 
light curves and we found smaller lags of few tens of seconds ($<$ 50 s) suggesting that 
the variation is probably originating from the corona.  
We found a broad noise component around $\sim$ 13 Hz in the 3-10 keV band which is absent in 10-20 keV band. We interpret the 
observed lags as the readjustment timescales of the corona or a boundary layer around 
the neutron star and constrain the height of this structure to few tens of km. 
We independently estimated the coronal height to be around 15 km assuming that 
the 13 Hz feature in the PDS is originating from the oscillation of the viscous shell 
around the neutron star. 
\end{abstract}



\section{Introduction}
Weakly magnetized Neutron Star Low mass X-ray binaries(NS LMXBs) can be 
broadly classified into Z and Atoll sources. Z sources, which are highly
X-ray luminous (0.5-1 L$_E$ : van der Klis 2005), are those that trace out a Z shaped track
in their Hardness Intensity Diagrams (HID) or Colour-Colour 
Diagrams (CCDs), while Atoll sources, emitting with lower X-ray
luminosity $\sim$ 0.001-0.5 L$_{Edd}$ 
(see e.g. Ford et al. 2000, van der Klis 2006 for a review), 
show two prominent branches, namely, the island and banana states (Hasinger \& van der Klis 1989; 
van der Klis 2006). Also, the island state can further be divided into upper
branch (UB) and lower branch (LB). NS LMXB in island state is found to have lower
luminosities and a harder spectra while in the banana state it is found to have
higher luminosities and a softer spectra (Barret 2001; Church et al. 2014). 
While the general idea has been that of the source traversing along the HID/CCD due 
to varying mass accretion rate, with $\dot{m}$ increasing 
from the island state towards the banana state, another 
scenario discusses the possibility of $\dot{m}$ remaining constant along
the track and the instabilities being caused by different accretion flow
solutions or radiation pressure at the inner radius or boundary layer or 
being the causative factor of the track traced in the HID (Homan et al. 2002, 2010;
Lin et al. 2007). Atoll sources are considered to have lower magnetic field and
accretion rate compared to Z sources. In the island state, it is hypothesized
that the disk is truncated relatively further away from the NS, while in the 
banana state the accretion disk approaches the NS (see Barret \& Olive 2002, Egron et al. 2013).

Although the truncated accretion disk model (Esin et al. 1997; see Done et al. 2007 for a review) and 
accretion disk-corona models (e.g., Church et al. 2006) propose geometrical configuration and evolution of LMXBs, 
questions still remain regarding the location and physical variability of 
 the coronal structure, which has remained more elusive for NS LMXBs compared to BH XRBs (Black Hole X-ray Binaries). 

Cross-Correlation Function (CCF) studies between softer and harder X-ray energy bands reveal important information
regarding the soft and hard X-ray emitting regions and the overall accretion disk-
corona geometry. The delay (or the relation in general) between soft and hard photons can be interpreted in 
the context of relative physical variations between the soft and hard X-ray emitting regions. 

Considering the soft photons, from say, the inner disk region to be reprocessed into
hard photons in the hotter Comptonization region, there must exist a few milli-second time lag/delay in the arrival of the
reprocessed photons. Such milli-second lags noticed in Z sources and BHXRBs are indicative of the 
Comptonization process in the corona/jet leading to hard lags (Vaughan et al. 1999; Kotov et al. 2001; Qu et al.
2001; Arevalo \& Uttley 2006; Reig \& Kylafis 2016). Soft milli-second lags where the soft photons lag the hard photons,
maybe explained by a shot model (Alpar \& Shaham 1985) or by a two-layer Comptonization model (Nobili et al. 2000).

Longer anticorrelated soft and hard delays, such as few tens to thousand second were initially interpreted as viscous timescales (readjustment
timescales) of the flow of matter in an optically thick accretion disk (Choudhary \& Rao 2004, Choudhary et al. 2005, Sriram et al. 2007, 2009, 2010)
based on CCF studies of BH XRBs. For the CCF study Choudhary et al. (2005) used 2-7 keV (soft) and 20-50 keV (hard) energy bands,
while Sriram et al. (2007, 2009, 2010) used 2-5 keV (soft) and 25-30 keV, 20-40 keV, 20-50 keV (hard) energy bands.
Here the geometrical picture used was that of a truncated accretion disk (soft photon emitting region)
with a hot compton cloud (source of hard photons) residing within the inner region of the disk, wherein any change in the disk structure will have to take place in viscous timescales with 
a corresponding change in the Compton cloud/corona. This disk readjustment timescale would therefore reflect the location of the truncation radius.
An accretion disk front moving toward the compact object (disk increasing in size) would result in more influx or increase of soft photons that 
would cool down the corona/Compton cloud and vice versa, causing variations in the geometry/size. 
Hence this association between the disk and the corona, 
was considered to be the cause of soft/hard delays.

But most recently, Sriram et al. (2019) performed CCF studies (2-5 keV and 16-30 keV) on the Z source GX 17+2, where few hundred 
second anticorrelated and correlated soft and hard time delays were detected when the source was in the HB and NB and even when the source was found to be 
close to the last stable orbit.  
Viscous timescale/readjustment timescale of the inner disk was estimated 
to be few tens of seconds, which falls short of what was observationally detected. 
Moreover lags were observed even on occasions when the estimated inner disk front (radius obtained from spectral fit) 
remained almost stationary. Hence the consideration that it is not just the 
disk but the readjustment of the entire disk-corona structure that leads to a few hundred second delays, resulted in the height of the coronal structure 
being constrained (see equation 2, Sriram et al. 2019). This was obtained by equating the observed CCF delay to the viscous timescale of the disk plus
that of the coronal structure, where the coronal velocity was assumed to be a fraction of the disk velocity ($\beta$ $\le$ 1). 
 Size of the coronal structure was constrained to be of the order of few tens of kilometres. Several other Z and atoll sources have also been found to exhibit delays of the same nature viz.
Cyg X-2 (Lei et al. 2008), GX 5-1 (Sriram et al. 2012), 4U 1735-44 (Lei et al. 2013),
4U 1608-52 (Wang et al. 2014) and it has been suggested that these lags could be readjustment in the inner region of the accretion disk. 
Lags were observed in GX 349+2 (Ding et al. 2016) where an extended accretion disc corona model was used for explaining them.

 A similar study on the nature of lags in atoll sources to constrain the coronal
size can help in throwing some light on the structural configuration of the coronal/sub-keplerian flow and 
maybe contribute towards resolving the dichotomy between atoll and Z sources.

4U 1705-44 is an atoll NS LMXB source (Hasinger \& van der Klis 1989). It is a type 1 X-ray burster
at a distance of 7.8 kpc (Galloway et al. 2008). It is a persistently bright source with an inclination
angle of 20$^\circ$ - 50$^\circ$ (Piraino et al. 2007).  Di Salvo et al. (2009) estimated the inner disk radius 
to be $\sim$ 2.3 ISCO, based on the reflection modeling of a relativistically smeared iron emission line identified
in the XMM-Newton spectral data of the source. 
While, from reflection spectral modeling of the iron line using Suzaku observations, Reis et al. (2009) estimated an inner disk 
radius of $\sim$ 1.75 ISCO. Cackett et al. (2010) 
estimated an inner disk radius of 1.0--6.5 ISCO by considering a blurred reflection model, 
with a blackbody component illuminating the disk in order to model the asymmetric Fe K emission lines in the Suzaku and XMM-Newton spectra.
The estimated inner disk radius was compared to that obtained from Diskbb model (continuum model) normalization parameter 
(R$_{in}$ (km) = $\sqrt{(N / cos i)}$ $\times$ D / 10 kpc)
and it was noted that R$_{in}$ from the continuum model is slightly lesser than that obtained from Fe line modeling.
Most recently, based on AstroSat LAXPC observations, 
Agrawal et al. (2018) found the inner disk radius to be 9--23 km ($<$ 2 ISCO) (before applying the spectral hardening/ colour correction 
factor)  based on DiskBB normalization. On applying the correction, the effective radius estimated was $\sim$ 26--68 km.
Although, the short coming of R$_{in}$ estimation from the Diskbb model would be the uncertainty involved due to the dependency
on inclination and source distance which are not well constrained quantities. 
Ford et al. (1998) and Olive et al. (2003) reported a low frequency noise around 10 Hz using RXTE observations.  
Agrawal et al. (2018) found a broad Peaked Noise (PN) component centered at 1-13 Hz, which was
attributed to the corona in the accretion disk.

The energy dependent CCF studies of 4U 1705–44 are
being reported here for the first time in LAXPC 3-5 keV vs.
15-30 keV energy bands in order to understand the accretion
disk corona geometry.

\section{Observations}
Regular pointing observations of 4U 1705-44, based on our proposal in AO cycle 3,
were made by AstroSat for 12 satellite orbits (Obs ID: A03\_073T01\_9000001498). Observations were
performed from 2017, August 29, 01:53:37 up to August 30, 01:06:48
for an effective exposure time of $\sim$ 36.8 ks (Dataset 1, see Table 1). Apart from this data, we have also used the AstroSat archival
data of 4U 1705-44 from March 2, 2017 to March 5, 2017 (Guaranteed Time (GT) phase) (Agrawal et al. 2018). This data
spans an exposure time duration of 100 ks (Dataset 2, see table 1).
For this work we have used the LAXPC (Large Area X-ray Proportional Counter)
data of the source. Data obtained in the Event Analysis (EA) mode has been used, which gives a time resolution of 10$\mu$s.

\begin{table*}
\tabletypesize{\scriptsize}
\caption{Details of datasets used for analysis.} 
\label{tab1}
\centering
\scalebox{0.85}{
\begin{tabular}{ccccccccc}
\hline
\hline
ObsID&&Orbit&&Start Time&&Stop Time\\
\hline
\hline
A03\_073T01\_9000001498 (Dataset 1)&&10375&&2017-08-29 01:53:37&&2017-08-29 02:27:33\\
"&&10376&&2017-08-29 02:03:33&&2017-08-29 04:10:17\\
"&&10377&&2017-08-29 03:31:39&&2017-08-29 05:55:03\\
"&&10378&&2017-08-29 05:18:27&&2017-08-29 07:37:15\\
"&&10379&&2017-08-29 07:01:41&&2017-08-29 09:21:29\\
"&&10380&&2017-08-29 08:42:28&&2017-08-29 11:05:56\\
"&&10381&&2017-08-29 10:52:25&&2017-08-29 12:51:19\\
"&&10382&&2017-08-29 12:34:15&&2017-08-29 14:38:52\\
"&&10383&&2017-08-29 14:25:58&&2017-08-29 16:22:09\\
"&&10386&&2017-08-29 15:57:33&&2017-08-29 21:36:13\\
"&&10387&&2017-08-29 21:15:20&&2017-08-29 23:17:44\\
"&&10389&&2017-08-29 22:38:47&&2017-08-30 01:06:48\\
G06\_064T01\_9000001066 (Dataset 2)	&&	7718	&&	2017-03-02 13:45:37	&&	2017-03-02 14:41:46	\\
"	&&	7719	&&	2017-03-02 14:18:40	&&	2017-03-02 16:26:30	\\
"	&&	7720	&&	2017-03-02 15:56:07	&&	2017-03-02 18:09:16	\\
"	&&	7722	&&	2017-03-02 17:33:46	&&	2017-03-02 19:27:04	\\
"	&&	7723	&&	2017-03-02 19:07:41	&&	2017-03-02 23:23:33	\\
"	&&	7724	&&	2017-03-02 22:43:44	&&	2017-03-03 01:06:58	\\
"	&&	7726	&&	2017-03-03 00:23:19	&&	2017-03-03 02:50:50	\\
"	&&	7727	&&	2017-03-03 02:26:18	&&	2017-03-03 04:39:58	\\
"	&&	7728	&&	2017-03-03 04:22:54	&&	2017-03-03 06:18:59	\\
"	&&	7729	&&	2017-03-03 06:01:29	&&	2017-03-03 08:02:35	\\
"	&&	7730	&&	2017-03-03 07:41:01	&&	2017-03-03 09:47:51	\\
"	&&	7731	&&	2017-03-03 09:36:58	&&	2017-03-03 11:35:55	\\
"	&&	7732	&&	2017-03-03 11:09:18	&&	2017-03-03 13:21:46	\\
"	&&	7733	&&	2017-03-03 12:52:45	&&	2017-03-03 15:04:49	\\
"	&&	7734	&&	2017-03-03 14:44:10	&&	2017-03-03 16:48:47	\\
"	&&	7737	&&	2017-03-03 16:17:07	&&	2017-03-03 22:03:14	\\
"	&&	7738	&&	2017-03-03 21:22:08	&&	2017-03-03 23:45:13	\\
"	&&	7740	&&	2017-03-03 23:23:43	&&	2017-03-04 01:29:44	\\
"	&&	7741	&&	2017-03-04 01:01:14	&&	2017-03-04 03:19:43	\\
"	&&	7742	&&	2017-03-04 03:05:52	&&	2017-03-04 04:56:56	\\
"	&&	7743	&&	2017-03-04 04:45:48	&&	2017-03-04 06:41:58	\\
"	&&	7744	&&	2017-03-04 06:26:40	&&	2017-03-04 08:26:15	\\
"	&&	7745	&&	2017-03-04 08:08:54	&&	2017-03-04 10:10:15	\\
"	&&	7746	&&	2017-03-04 09:52:44	&&	2017-03-04 12:01:17	\\
"	&&	7747	&&	2017-03-04 11:35:23	&&	2017-03-04 13:44:45	\\
"	&&	7748	&&	2017-03-04 13:16:25	&&	2017-03-04 15:27:58	\\
"	&&	7749	&&	2017-03-04 14:53:49	&&	2017-03-04 17:12:44	\\
"	&&	7751	&&	2017-03-04 16:49:47	&&	2017-03-04 18:44:10	\\
"	&&	7752	&&	2017-03-04 18:43:21	&&	2017-03-04 22:23:57	\\
"	&&	7753	&&	2017-03-04 21:58:07	&&	2017-03-05 00:07:27	\\
"	&&	7755	&&	2017-03-04 23:33:10	&&	2017-03-05 01:52:42	\\
"	&&	7756	&&	2017-03-05 01:15:47	&&	2017-03-05 03:42:19	\\
"	&&	7757	&&	2017-03-05 03:21:18	&&	2017-03-05 05:21:04	\\
"	&&	7759	&&	2017-03-05 05:15:32	&&	2017-03-05 08:51:08	\\

\hline
\end{tabular}}
\end{table*}

LAXPC has three co-aligned identical proportional counters viz. LAXPC 10, 20 and 30 each
with a total effective area of 6000 cm$^{2}$ at 15 keV. It operates in 
the 3-80 keV energy range (Yadav et al. 2016a; Antia et al. 2017). It has a field of
view of 1$^\circ$ $\times$ 1 $^\circ$ and a moderate energy resolution of 15 \%, 12 \% and 
11 \% at 30 keV for LAXPC 10, 20 and 30 respectively (Yadav et al. 2016; Antia et al. 2017).

\section{Data Reduction and Analysis}
For reducing the Event Analysis (EA) mode LAXPC level 1 data, we
have used the LAXPC software (format A, 2018 May 19 version) provided by the AstroSat Science Support Center
(ASSC). Following the standard procedures of \textsc{\lowercase{LAXPC\_MAKE\_EVENT}} and 
\textsc{\lowercase{LAXPC\_MAKE\_STDGTI}}, the event file and good time interval file (GTI), with
the Earth occultation and South Atlantic Anomaly (SAA) removed,
was generated. The event file and GTI file were used to generate the
light curves using the standard routine of \textsc{\lowercase{LAXPC\_MAKE\_LIGHTCURVE}}.
LAXPC 10 data was used for producing the light curves for the timing
analysis.

\section{Timing Analysis} 
Lightcurves from LAXPC 10 were generated for timing analysis.
 Figure 1 shows the 32 s binned background subtracted lightcurves in the 3-30 keV energy
range. Top panel shows the light curve obtained from dataset 1 which spans $\sim$ 36.8 ks and
the bottom panel shows the same for dataset 2 which spans $\sim$ 100 ks.

For obtaining the HID for dataset 1, lightcurves were generated in the energy bands 
viz. 3.0-18.0 keV, 7.5-10.5 keV and 10.5-18.0 keV  energy range (Agrawal et al. 2018),
where hardness ratio is that between the count rates in 7.5-10.5 keV and 
10.5-18.0 keV energy bands and intensity is that in the 3.0-18.0 keV energy range (Fig. 2, For more
details see Malu et al. (2021)).
Based on the HID of both datasets, 4U 1705-44 is found to be in the banana state.
The HID was divided into 3 parts for Dataset 1 - A, B and C sections in order to study the energy-rms relation,
Power Density Spectra (PDS) and CCFs. Figure 3 shows that energy-rms plot for light curves of the three sections - A, B and C.
Light curves where obtained in the 3-5 keV, 5-7 keV, 7-9 keV, 9-11 keV, 11-13 keV,
13-15 keV, 15-20 keV, 20-25 keV and 25-30 keV in order to determine the fractional rms variation in the light curve.
The trend is found to be increasing in all the three cases, as often seen in these sources. Right panel of the figure gives the 
PDS of the three sections respectively in the 3-30 keV energy range using a lightcurve of 1/512 s binsize (described later in detail).

\begin{figure}[!t]
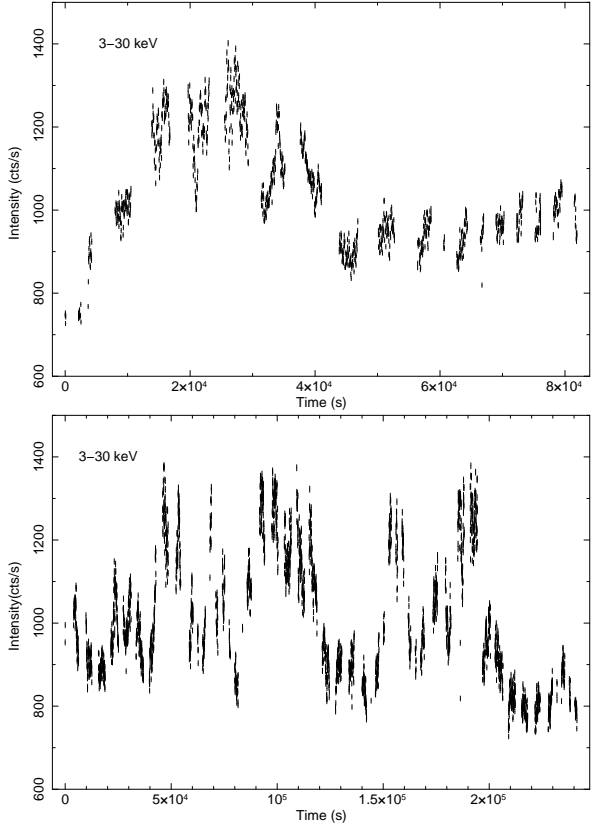

\includegraphics[width=.65\columnwidth, angle=270]{lc.ps}
\includegraphics[width=.65\columnwidth, angle=270]{lc_dataset2.ps}
\caption{Top: Lightcurve (dataset 1) of 4U 1705-44 obtained from LAXPC 10 binned with 32 s binsize in the 3-30 keV energy band. 
Bottom: Same for dataset 2.}\label{fig1}
\end{figure}

\begin{figure}[!t]
\includegraphics[width=.65\columnwidth, angle=270]{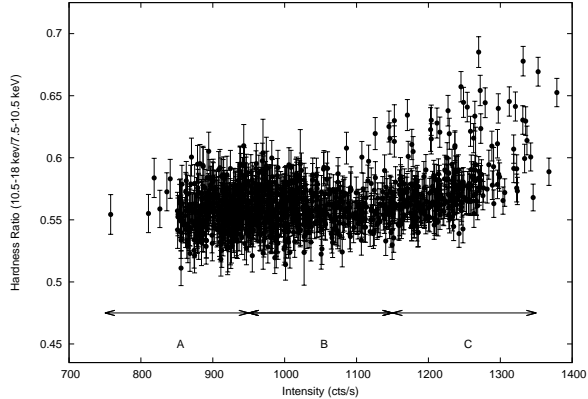}
\caption{HID for 4U 1705-44 using AstroSat LAXPC observations (Dataset 1). Hard colour is defined as 10.5-18/7.5-10.5 keV. Intensity
is given in the 3-18 keV energy range. Arrows mark the separation of HID into A, B and C sections (see text).}\label{fig2}
\end{figure}

\begin{figure*}[!t]
\includegraphics[width=10.0cm,height=8.0cm, angle=270]{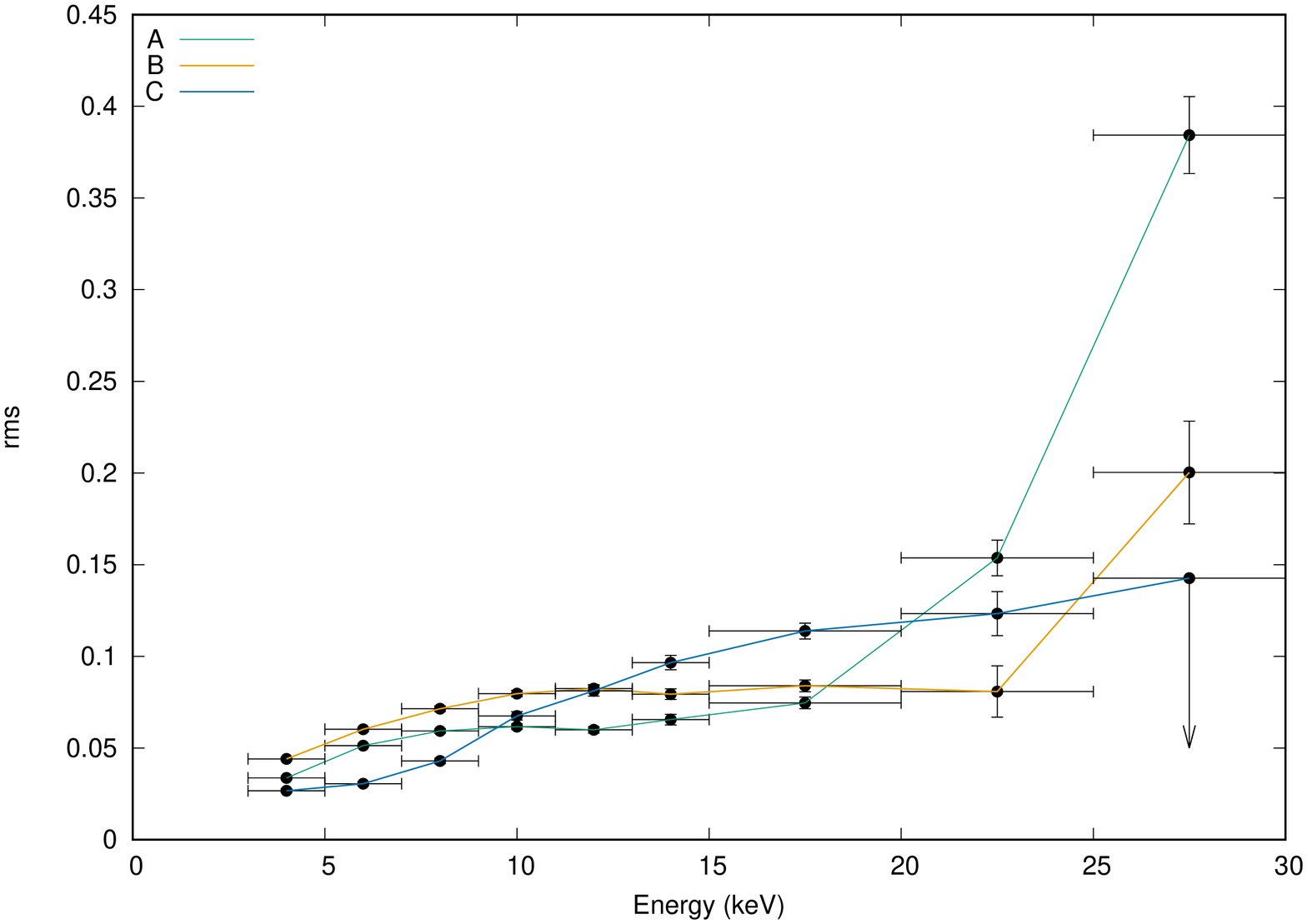}
\includegraphics[width=10.0cm,height=8.0cm, angle=270]{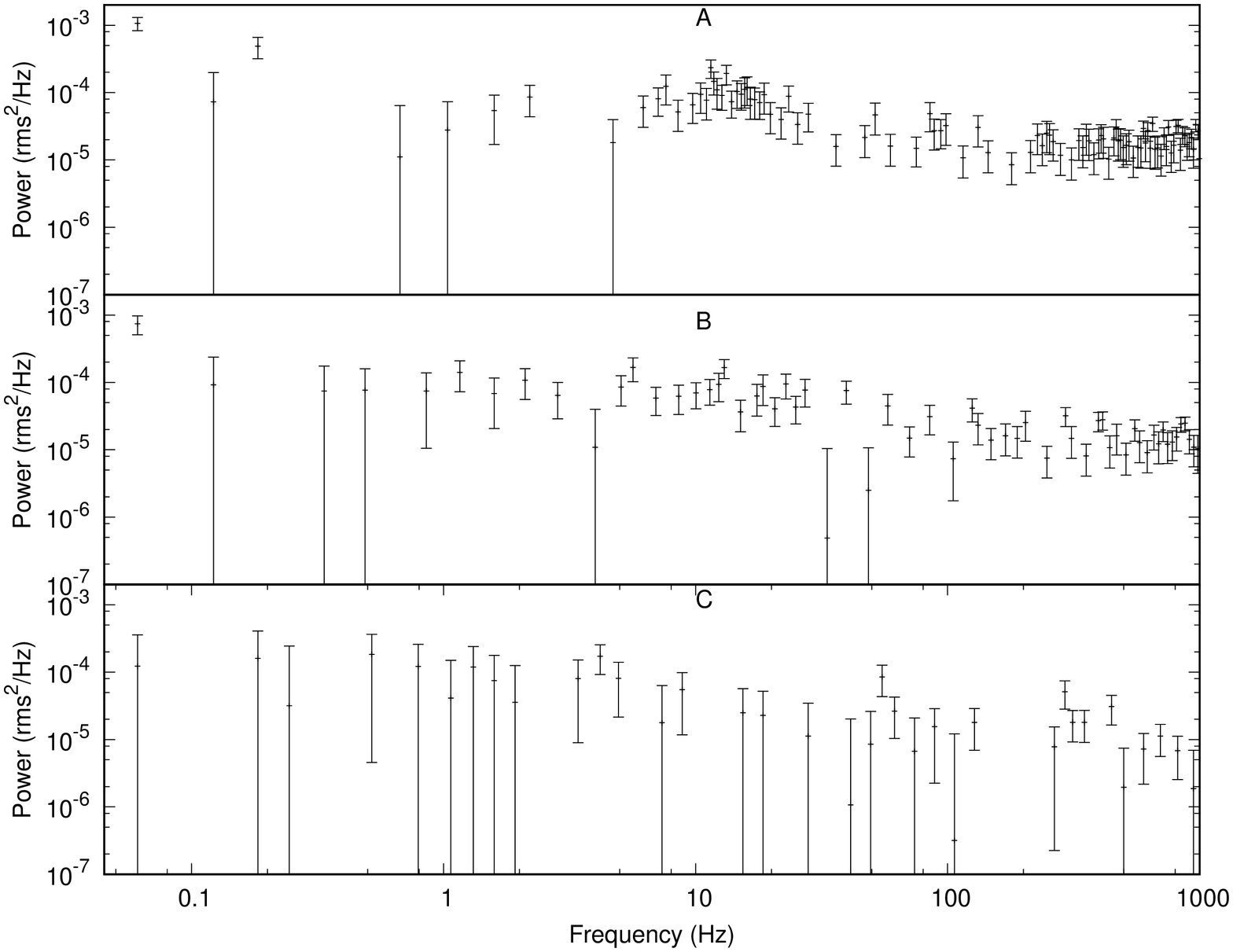}
\vspace{1.5cm}
\caption{Left: Energy-rms relation for each section of the HID marked A, B and C. The arrow-mark indicates the 
3 $\sigma$ upper limit. Right: PDS for the A,B and C sections.}\label{fig3}
\end{figure*}

\subsection{Cross-Correlation Function study}
Lightcurves in the 3-5 keV (soft) and 15-30 keV (hard) energy bands
were extracted for performing CCF studies. The
 \textsc{\lowercase{CROSSCOR}} tool available in the XRONOS package was utilized for performing
the cross correlation analysis (see Sriram et al. 2007, 2011, 2012, 2019, 
Lei et al. 2008, 2013 and Malu et al. 2020). The \textsc{\lowercase{CROSSCOR}} tool uses a FFT or
a direct slow (option fast=no) algorithm to compute cross-correlation between
two simultaneous light curves and gives an output of CC value as a function of time delay.
We have used the slow direct mode and in this mode the CCF error bars are obtained 
by propagating the  theoretical error bars of the  cross  correlations from  individual
intervals that are in  turn  obtained by propagating the newbin error bars
through the cross  correlation  formula\footnote{https://heasarc.gsfc.nasa.gov/docs/xanadu/xronos/help/crosscor.html}.
These cross-correlations are then normalized by dividing them by the square root of the
product of the number of good newbins of the two time series in each interval, to obtain
the cross covariances.

A 10 s bin size was used during the CCF analysis and lags were obtained in ten different light curve
segments of the order of hundred seconds (Fig. 4, 5). Gaussian function was fitted on the data points around the
most significant peak in the CCF profile with a 90\% confidence level (Fig. 6). A $\chi$$^2$ minimization method with the criterion of 
$\Delta$$\chi$$^2$ = 2.7, was used in order to determine the CCF lag values and the corresponding errors (Table 2).

Asymmetric CCF profiles are difficult to judge and they clearly can not be approximated
by a Gaussian profile. We have used the Gaussian profile fitting method only on those CCF profiles
that are relatively symmetric and could be approximated to a Gaussian profile. For this we have set a criterion
of $\chi$$^2$/dof $<$ 2. Even so, the lags
thus estimated should be cautiously taken. For further results we have only considered those lag
values that were best constrained with lower error bars following our set criterion. Larger uncertainties in the CCF lag estimation
would directly lead to larger disparities in the results further discussed. The $\chi$$^2$/dof values would
depend upon the number of points around the peak CC that are considered for the fit. Errors and lag
values would depend upon this consideration eventually.

For dataset 1, amongst the observation during 12 orbits, three lightcurve segments exhibited lags. 
Figure 4a-c shows the lightcurve segments and the corresponding CCFs of each segment. 
Segments that showed delays had symmetric profiles and low error
bars. Hence, all the three sections that showed lags were considered for further studies. 
While one light curve segment showed anti-correlated hard lags of 371 $\pm$ 7 s (CC = -0.53 $\pm$ 0.06),
one segment showed an anti-correlated 
soft lag of 163 $\pm$ 9 s (CC = -0.47 $\pm$ 0.08) (see fig. 4a-c. Table 2). CCF in Fig. 4c. (top panel)
peaks positively at 0 and negatively at $\sim$ 258 s (hard lag) both with similar CC values, 
hence in such a case specifying the exact nature and occurrence of the lag is rather tough.

By hard lag we are referring to the hard photons lagging the soft 
photons and the opposite scenario for the soft lags. Most of the remaining segments
were positively correlated with no lag (CC $\sim$ 0.6-0.8) while few other segments
were uncorrelated. All the lags noted lie in section A and B of the HID.

\begin{table*}
\begin{minipage}[ht]{\columnwidth}
\scriptsize
\caption{Obtained CCF lags in dataset 1 and 2 with their corresponding errors (90 \% confidence limits) and the best fit $\chi$$^{2}$/dof.} 
\label{tab1}
\centering
\begin{tabular}{ccccccccc}
\hline
\hline
&Dataset 1&&&\\
\hline
CCF lag$\pm$error (s) &CC $\pm$ error&$\chi$$^{2}$/dof&Corresponding Figures\\
\hline
371 $\pm$ 7&-0.53 $\pm$ 0.06&46.13/66 & Fig. 4a (bottom panel)\\
-163 $\pm$ 9&-0.47 $\pm$ 0.08&22.24/29 & Fig. 4b (bottom panel)\\

\hline
&Dataset 2 &&&\\
\hline
64 $\pm$ 6&0.53 $\pm$ 0.10&22.64/27& Fig. 5a (bottom panel)\\
-274 $\pm$ 21 &-0.27 $\pm$ 0.09&52.05/77 & Fig. 5b\\
-137 $\pm$ 14 &-0.28 $\pm$ 0.05&146.3/109 & Fig. 5c (top panel)\\
-533 $\pm$ 8 & -0.50 $\pm$ 0.06 &69.92/109 & Fig. 5d (bottom panel)\\
573 $\pm$ 28&-0.31 $\pm$ 0.05&391.8/276 &  Fig. 5e\\
-450 $\pm$ 21 & 0.31 $\pm$ 0.05& 140.9/197 & Fig. 5f\\
238 $\pm$ 15 &-0.31 $\pm$ 0.10&47.64/70 & Fig. 5g (bottom panel)\\

\hline
\hline
\hline
\end{tabular}
\end{minipage}
\end{table*}

\begin{subfigures}

\begin{figure*}[!ht]
\caption{Dataset 1: The background subtracted 10 s bin LAXPC soft (3--5 keV) and hard X-ray (15--30 keV) light curves (left panels) for which CCF lags are observed (right panels).
Energy bands used are mentioned in the light curves (left panel). Right panels show the cross correlation function (CCF) of each section of the light curve and shaded regions show the standard deviation 
of the CCFs. The vertical line on the light curves in the left panel separates the two sections for which CCF lags are shown on the right panel.}\label{fig4}
\includegraphics[width=10.0cm,height=18.cm, angle=270]{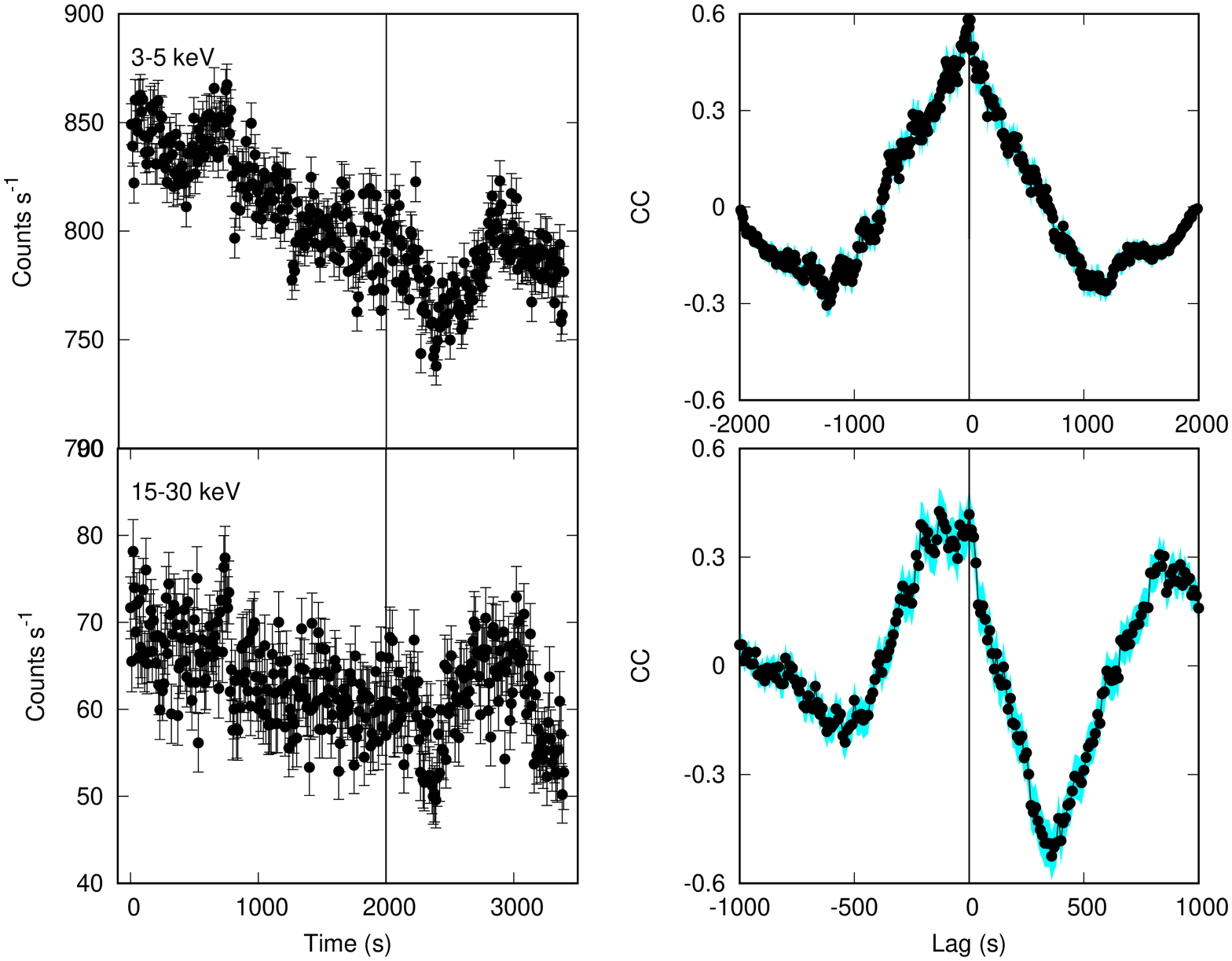} \\
\end{figure*}
\begin{figure*}
\caption{(b)}
\includegraphics[width=10.0cm,height=18.cm, angle=270]{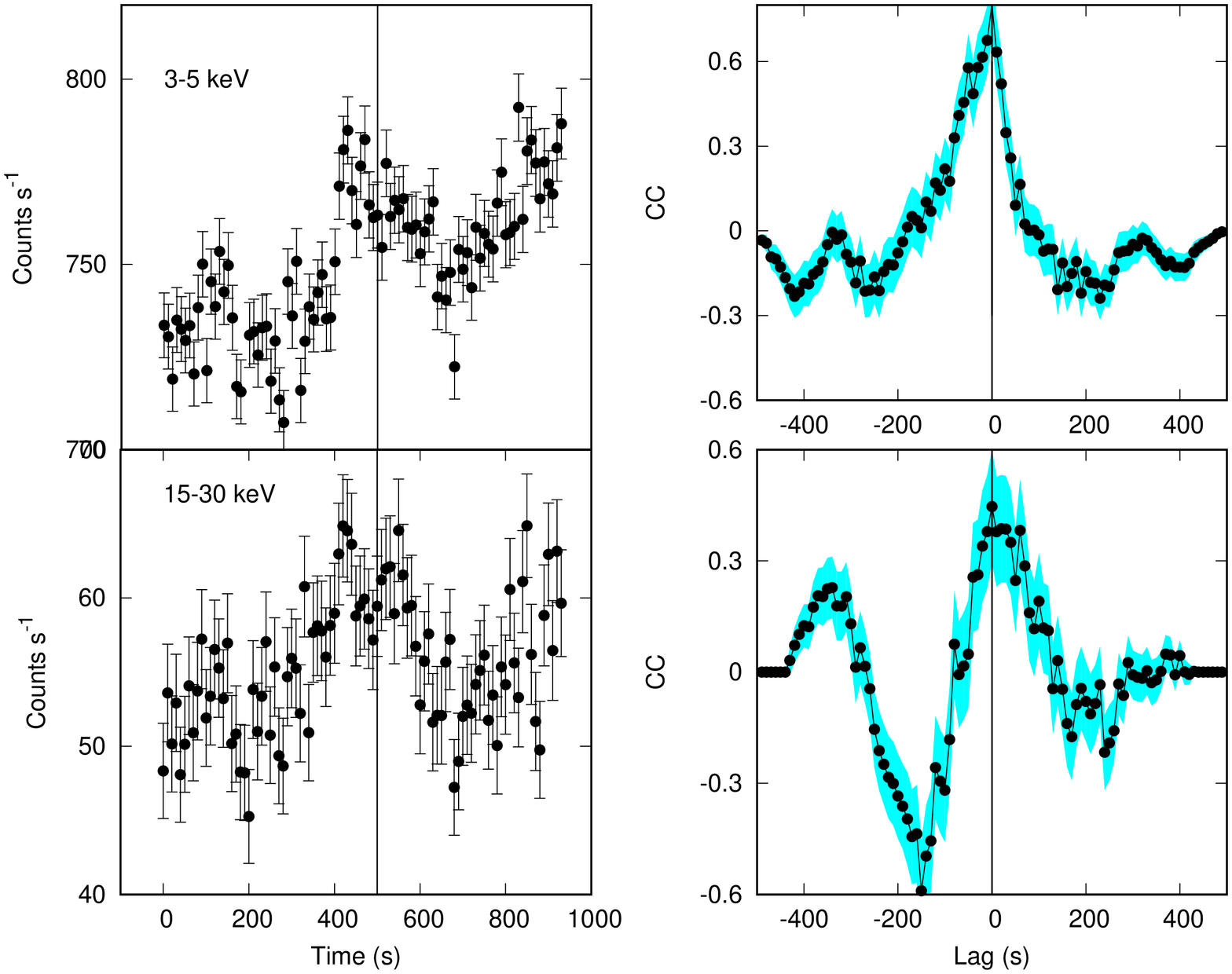} 
\end{figure*}
\begin{figure*}
\caption{(c)}
\includegraphics[width=10.0cm,height=18.cm, angle=270]{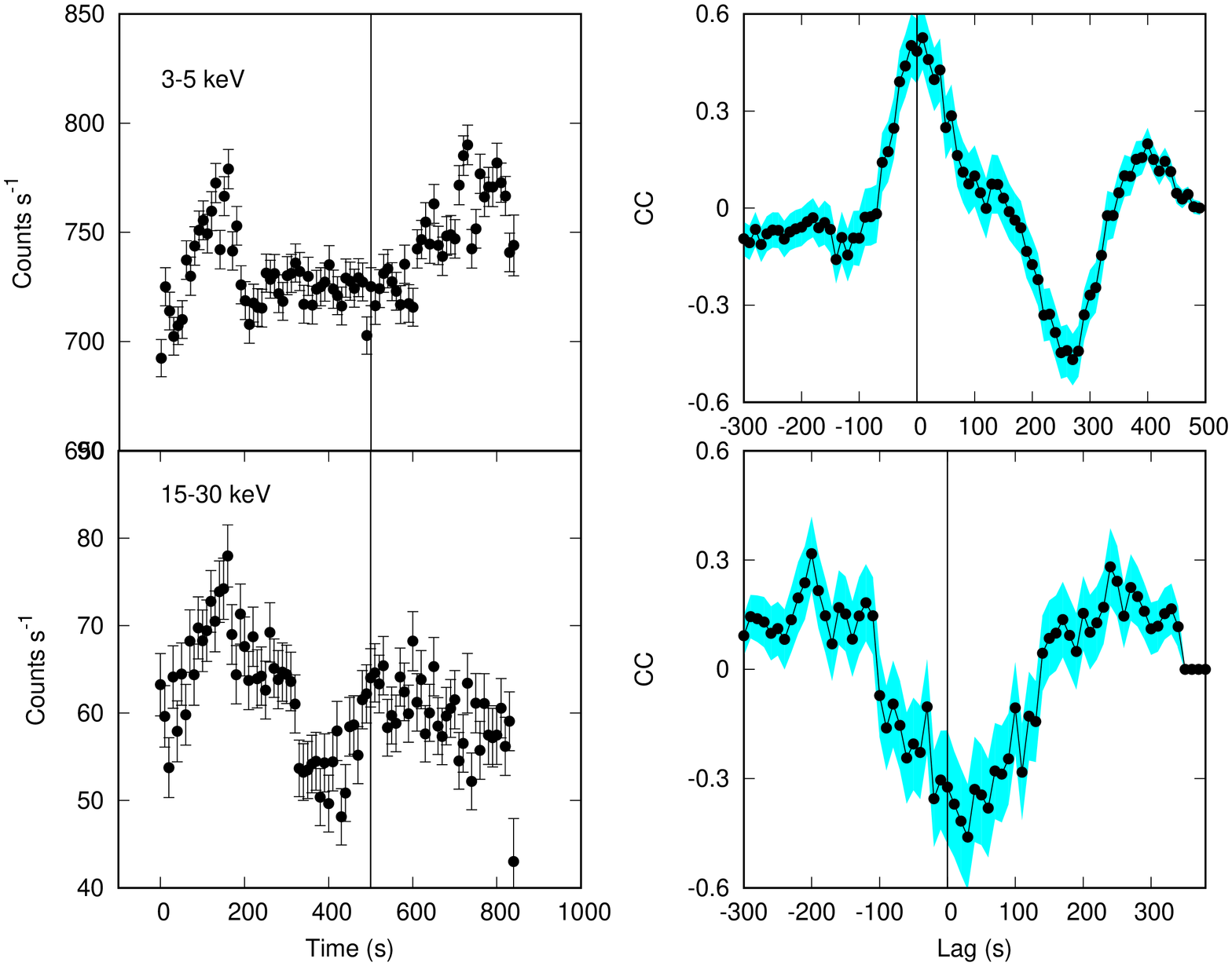} \\
\end{figure*}
\end{subfigures}

For dataset 2, amongst the observation during 34 orbits, seven light curve segments showed signatures of lags.
Three sections out of these were ruled out based on asymmetry, 
of which one showed significant error. Hard/soft anticorrelated/correlated lags were noted viz.
64 $\pm$ 6 s (CC = 0.53 $\pm$ 0.10), -274 $\pm$ 21 s (CC = -0.27 $\pm$ 0.09), -137 $\pm$ 14 s (CC = -0.28 $\pm$ 0.05),
-533 $\pm$ 8 s (CC = -0.50 $\pm$ 0.06), 573 $\pm$ 28 s (CC = -0.31 $\pm$ 0.05), -450 $\pm$ 21 s (CC = 0.31 $\pm$ 0.05)
and 238 $\pm$ 15 s (CC = -0.31 $\pm$ 0.10) (see fig 5 a-g, Table 2). Most of the remaining light curve segments were positively correlated
with CC $\sim$ 0.6-0.8 and few others were uncorrelated.

\begin{subfigures}
\begin{figure*}[!ht]
\caption{Dataset 2: The background subtracted 10s bin LAXPC soft (3-5 keV) and hard X-ray (15--30 keV) light curves (left panels) for which CCF lags are observed (right panels).
Energy bands used are mentioned in the light curves (left panel). Right panels show the cross correlation function (CCF) of each section of the light curve and shaded regions show the standard deviation 
of the CCFs. The vertical line on the light curves in the left panel separates the two sections for which CCF lags are shown on the right panel.}\label{fig5}
\includegraphics[width=10.0cm,height=18.cm, angle=270]{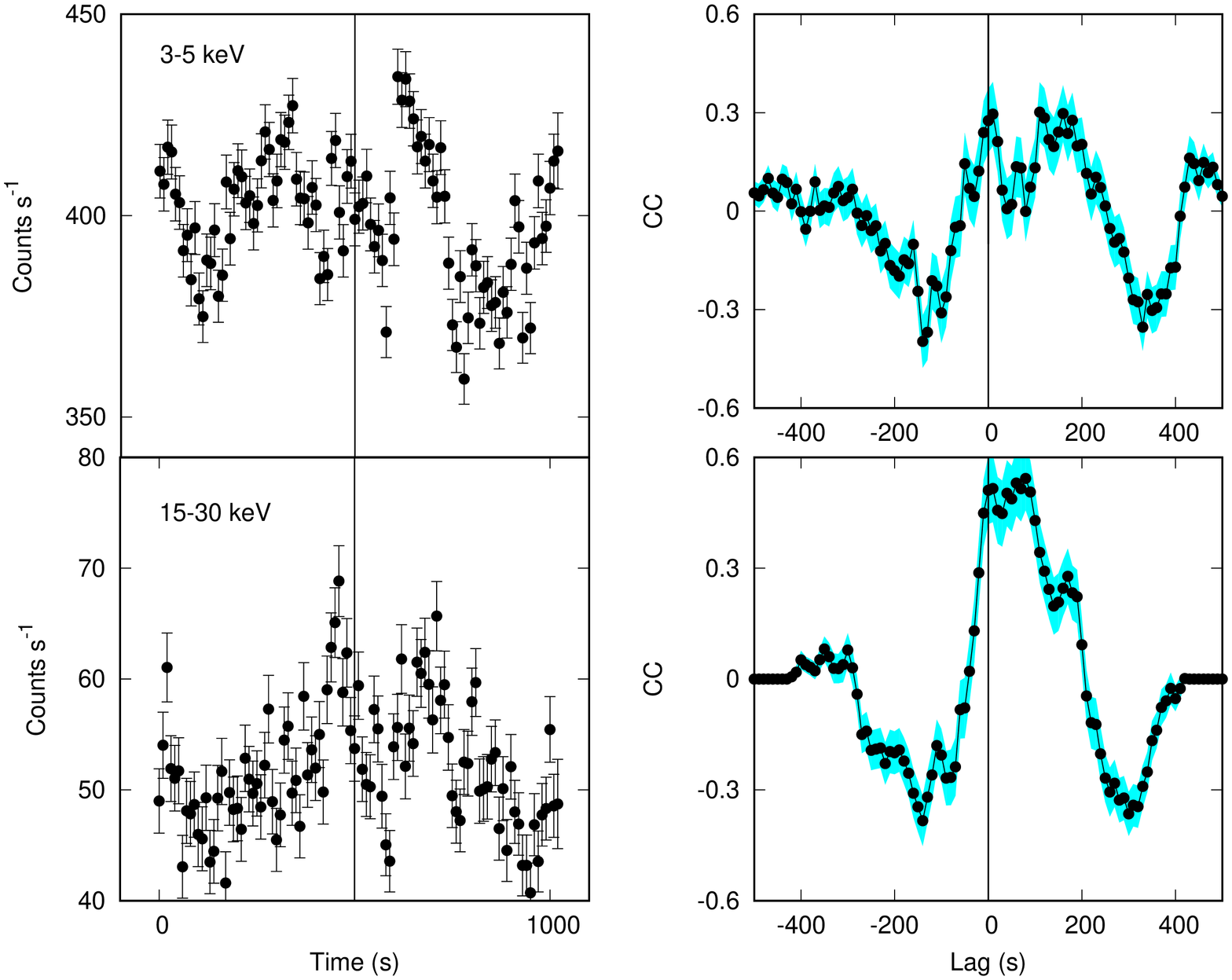} \\
\end{figure*}
\begin{figure*}
\caption{(b)}
\includegraphics[width=10.0cm,height=18.cm, angle=270]{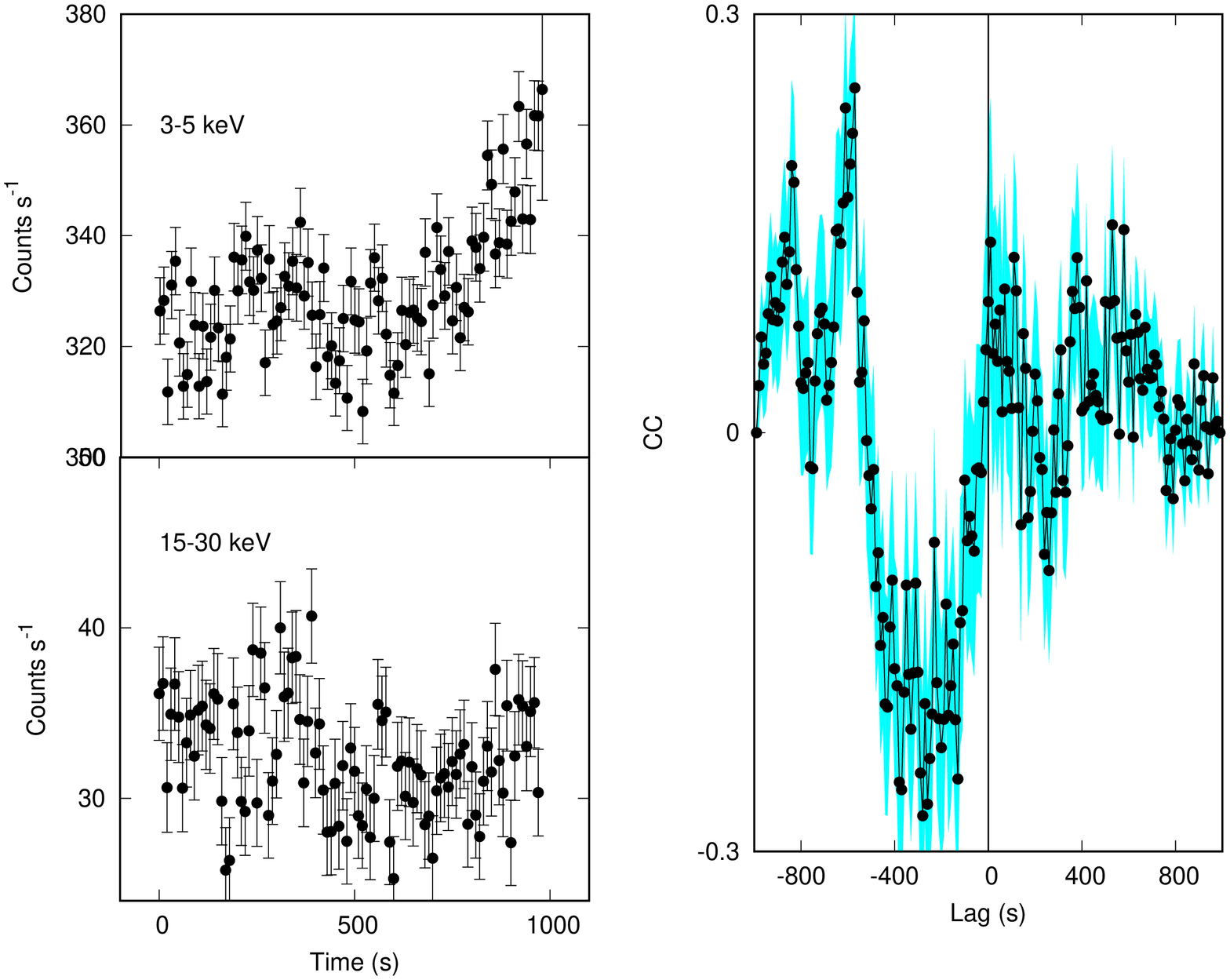} \\
\end{figure*}
\begin{figure*}
\caption{(c)}
\includegraphics[width=10.0cm,height=18.cm, angle=270]{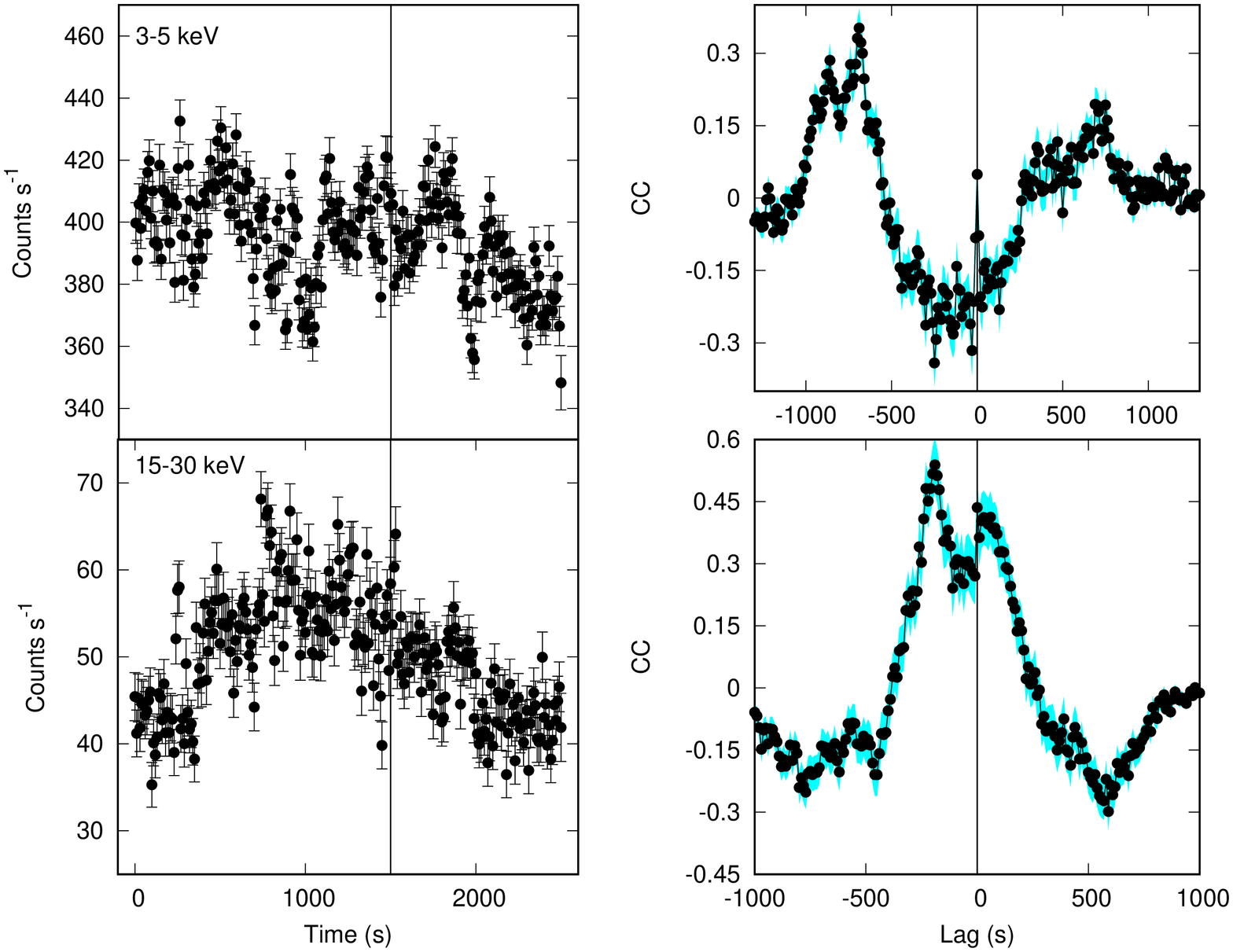} \\
\end{figure*}
\begin{figure*}
\caption{(d)}
\includegraphics[width=10.0cm,height=18.cm, angle=270]{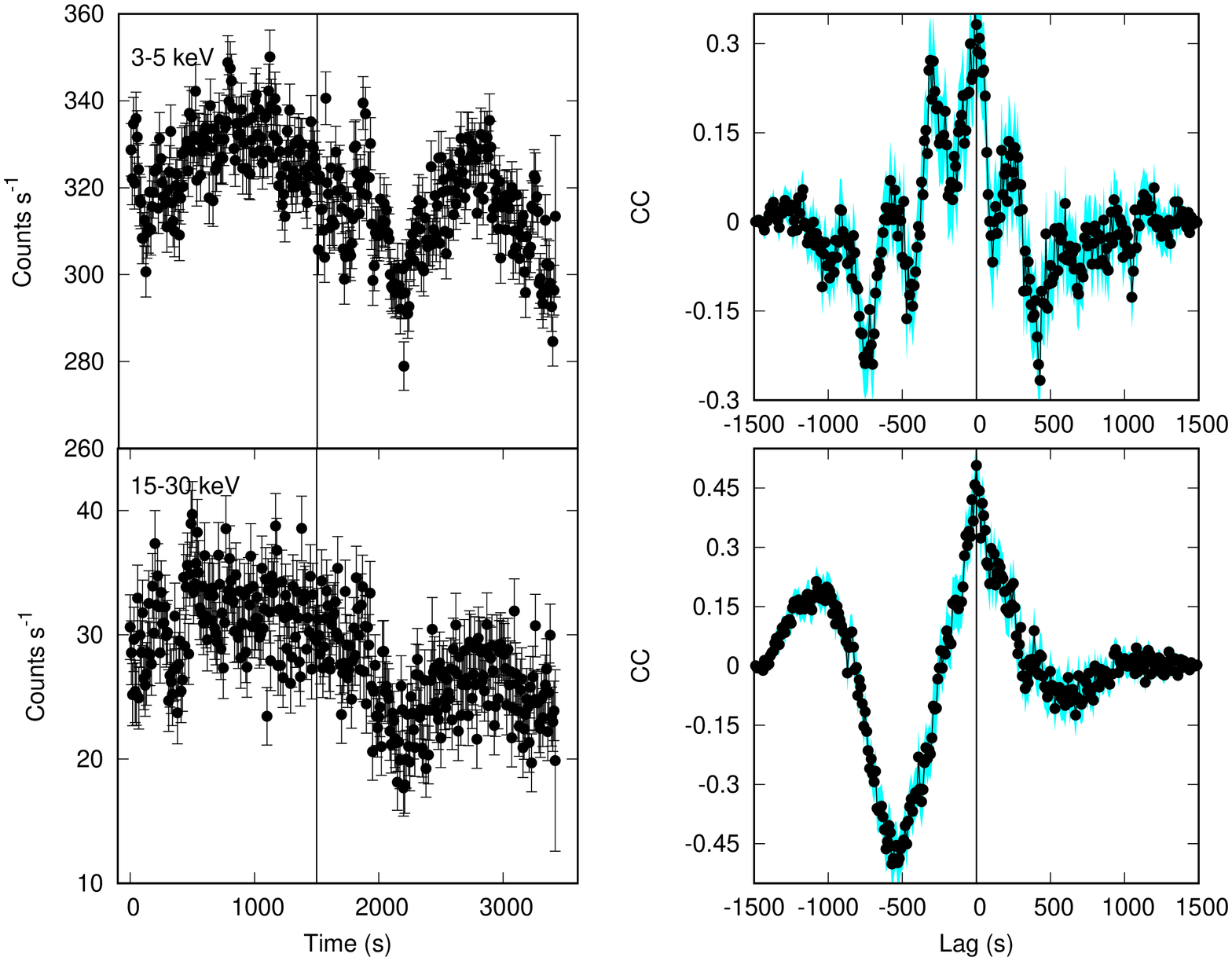} \\
\end{figure*}
\begin{figure*}
\caption{(e)}
\includegraphics[width=10.0cm,height=18.cm, angle=270]{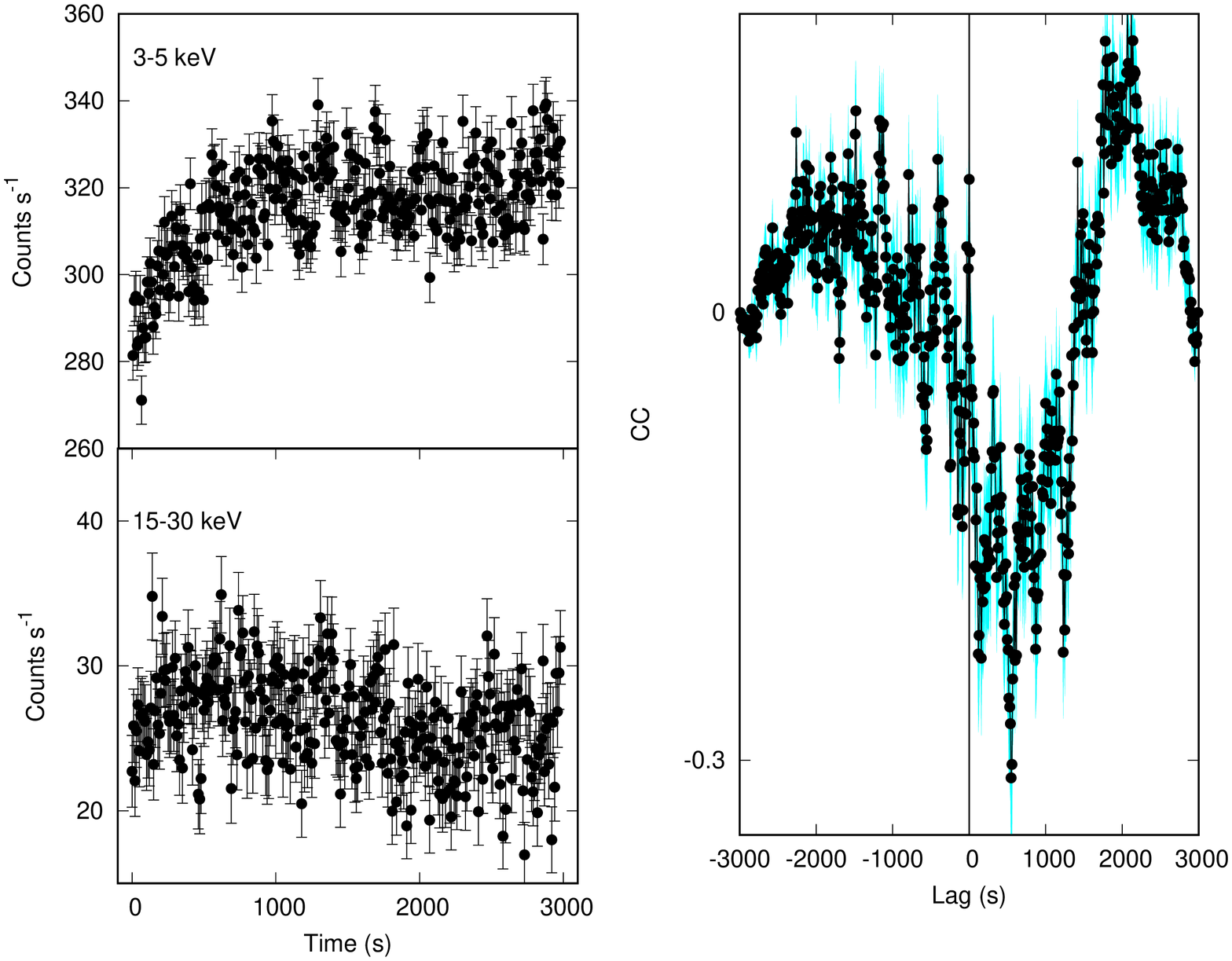} \\
\end{figure*}
\begin{figure*}
\caption{(f)}
\includegraphics[width=10.0cm,height=18.cm, angle=270]{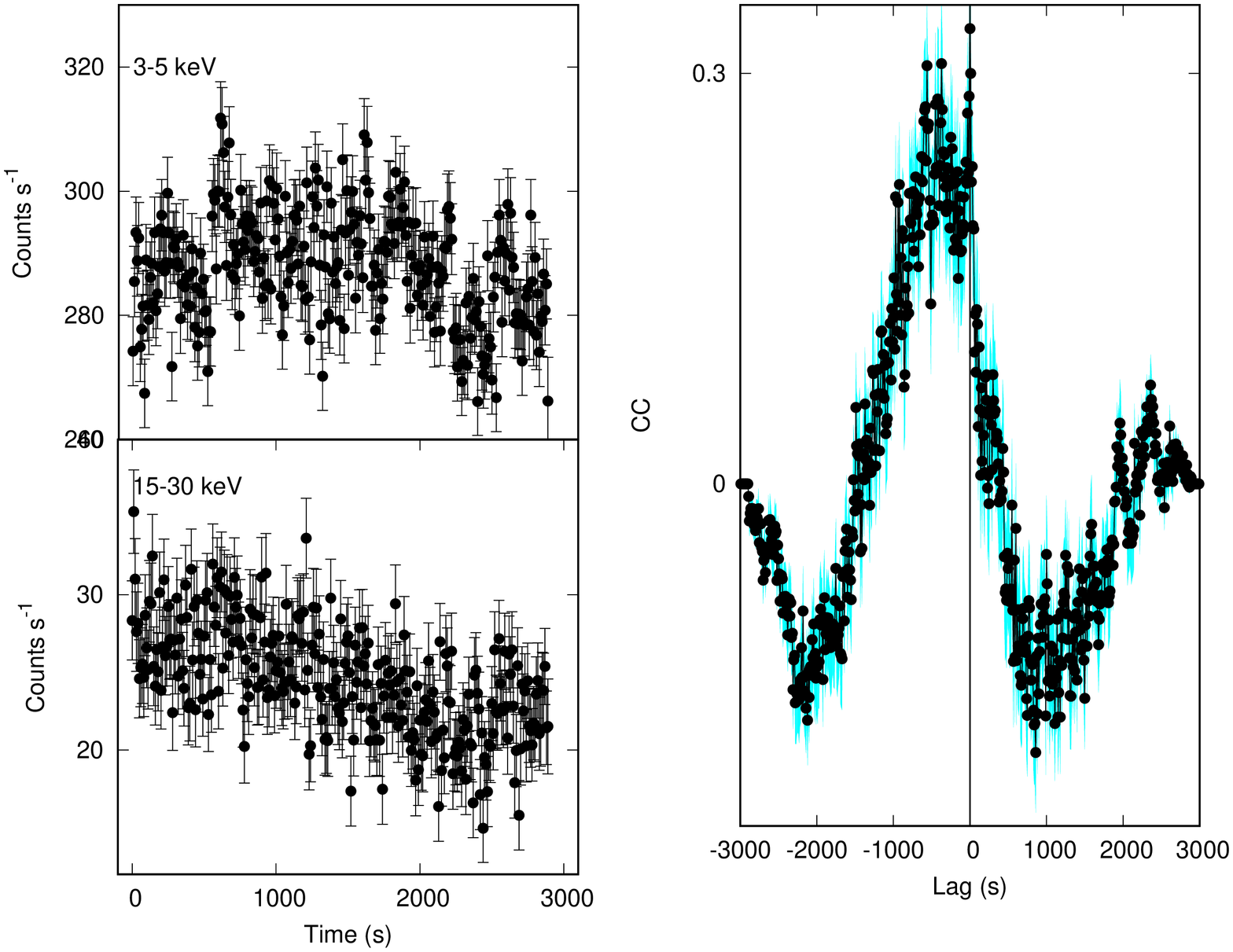} \\
\end{figure*}
\begin{figure*}
\caption{(g)}
\includegraphics[width=10.0cm,height=18.cm, angle=270]{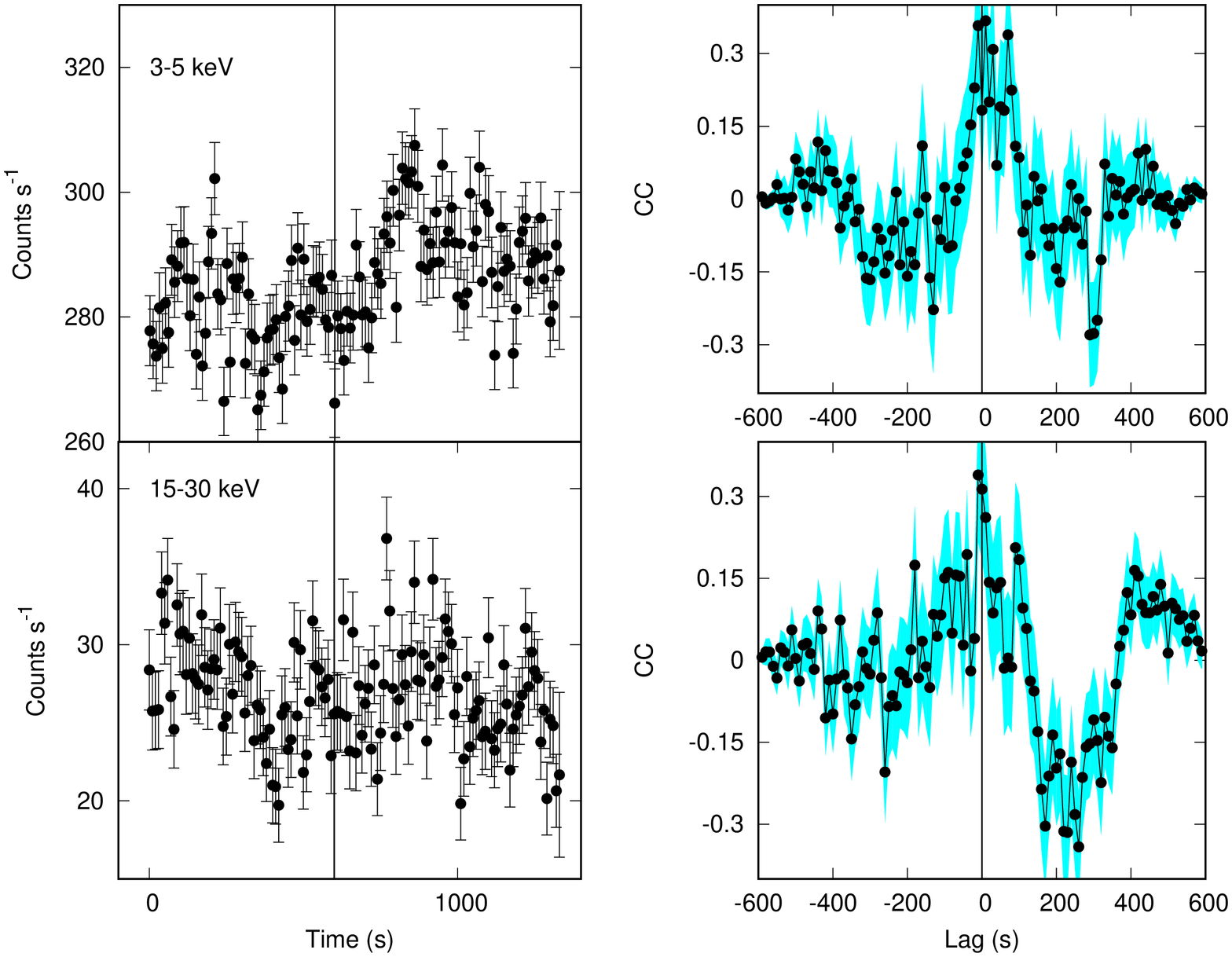} \\
\end{figure*}
\end{subfigures}

\begin{figure*}[!ht]
\includegraphics[width=10.0cm,height=18.0cm, angle=270]{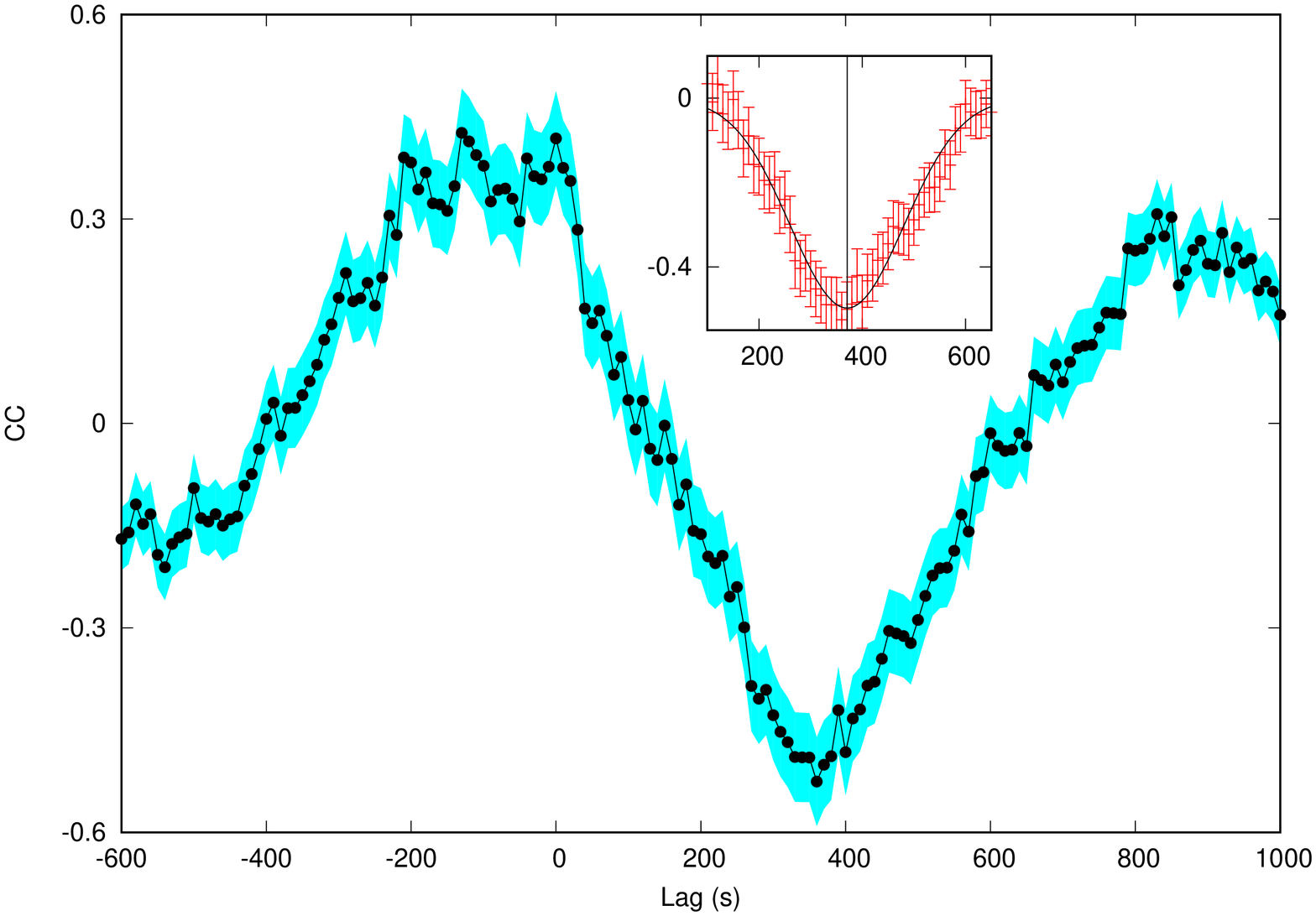}
\caption{Figure shows a representative CCF lag fitted using a Gaussian function using points around the maximum peak 
lag = 371 $\pm$ 7, CC value = -0.53 $\pm$ 0.06, $\chi$$^2$/dof = 46.13/66.}\label{fig6}
\end{figure*}

 The complex structure and evolution of CCF structures or the energy dependencies of light
curves on shorter Fourier time scales can be probed by performing a dynamic CCF study 
that uses shorter correlation time scales. Dynamic CCF study between 3-5 keV and 15-30 keV energy bands was 
performed on light curve segments of size 100 s each with a bin size of 10 s.
CCF lags of the order of few tens of seconds were clearly noticed in the dynamical CCF study 
between 3-5 keV and 15-30 keV (see fig. 7), as can be noted in the left panel.
The middle panel in figure 7 shows few noticeable CCFs that were obtained during the study.
The broad CCF around zero lag indicates that the correlation is weak. 
The correlation strength of the CCFs i. e. CC coefficients are depicted using different colours with
values +1 to -1 showing maximum positive correlation and maximum anticorrelation respectively.  
Energy dependent light curves used in this study are shown in the right panel.

\begin{figure*}[!t]
\centering

\includegraphics[width=1.25\columnwidth, angle=270]{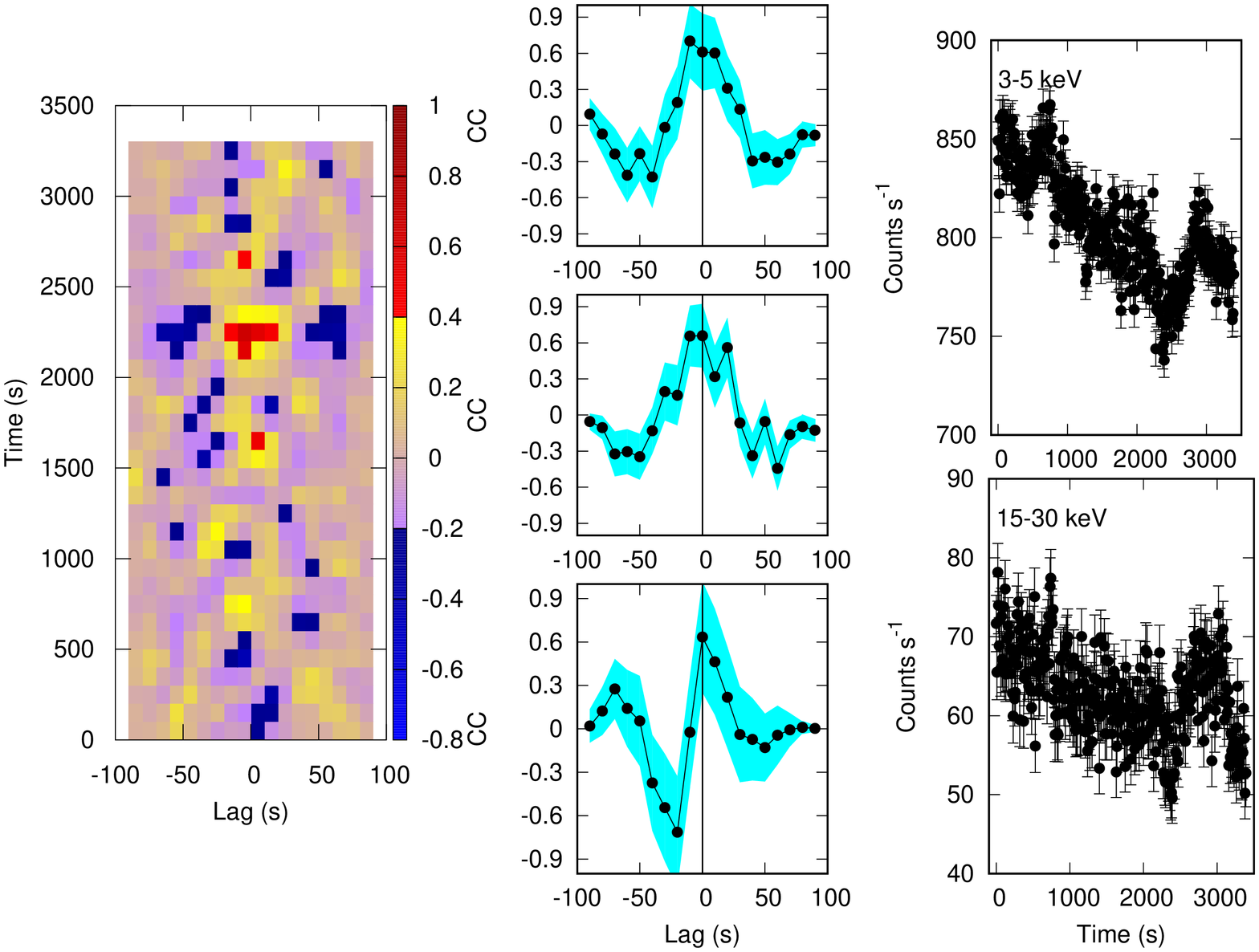} 
\vspace{2cm}
\caption{Left panel shows the dynamical
CCF between soft (3-5 keV) and hard (15-30 keV) energy
bands. Dynamical CCF has been obtained using a bin size of 10 s for light curve segments of size 100 s each. 
Hence CCF delays (or lags) occurring within 100s  can be noted distinctly. Colour scale depicts the
strength of the CC coefficients/values at different lags where +1 represents maximum positive correlation and -1 
represents the maximum anticorrelation. Middle panel shows few noticeable representative CCFs for this light curve segment.
The corresponding energy dependent light curves are shown in the right panel.}\label{fig7}
\end{figure*}

Power Density Spectrum (PDS) of section A, B and C was obtained (dataset 1) using a lightcurve of 1/2048 s binsize
in the energy range 3-30 keV. Dead time corrected Poisson noise level was subtracted from all the PDS based on Yadav et al. (2016b).
Only the PDS of section A revealed a broad feature around 5-13.5 Hz, similar to the PDS reported by Agrawal et al. (2018) for dataset 2. 
Each orbit's lightcurve was independently searched for this broad Peaked Noise component and
we noted this feature centered around $\sim$ 5-13.5 Hz in only a few of the 
light curve segments (all lying in A and B sections of the HID). 
The most prominent feature had a quality factor (ratio of the QPO central frequency to its full width at half maximum), 
Q $\sim$ 2.14 (centroid frequency 13.2 $\pm$ 0.6 Hz).
A similar search for this feature was made separately in the 3-10 keV and 10-20 keV energy bands
in order to constrain its origin (see Table 3). It was found that the feature exists only 
in the 3-10 keV energy band with rms amplitude of 4.13 \% (similar to that noted by Agrawal et al. 2018) and not in the 10-20 keV band 
where an upper limit of $\sim$ 0.66 \% was estimated (See fig. 8, Table 3). 

\begin{table}
\begin{minipage}[ht]{\columnwidth}
\scriptsize
\caption{Best fit parameters for PDS in the 3-10 keV energy range (detection) and 10-20 keV energy range  (non detection).} 
\label{tab3}
\centering
\begin{tabular}{ccccccccc}
\hline
\hline
&3-10 keV&10-20 keV\\
\hline
&PN&NO PN\\
\hline
$\Gamma_{pl}$\footnote{Power-law index.}& -0.71$\pm$0.03&--\\
N$_{pl}$ ($\times$ 10$^{-4}$) \footnote{Power-law Normalization}&2.89 $\pm$0.23&--\\
PN $\nu$\footnote{centroid frequency (Hz)}&  13.20 $\pm$0.62&--\\
FWHM (Hz) & 6.16 $\pm$ 1.87&--\\
N$_{l}$ ($\times$ 10$^{-4}$) \footnote{Lorentzian Normalization }&1.59 $\pm$0.31&--\\

rms \% & 4.13 \%&  $<$ 0.66\% (upper limit)\\ 
Q-factor & 2.14 &--\\
$\chi$$^{2}$/dof&  155/116 &--\\
\hline
\hline
\hline
\end{tabular}
\end{minipage}
\end{table}

\begin{figure}[!ht]
\includegraphics[width=10.0cm,height=8.0cm, angle=270]{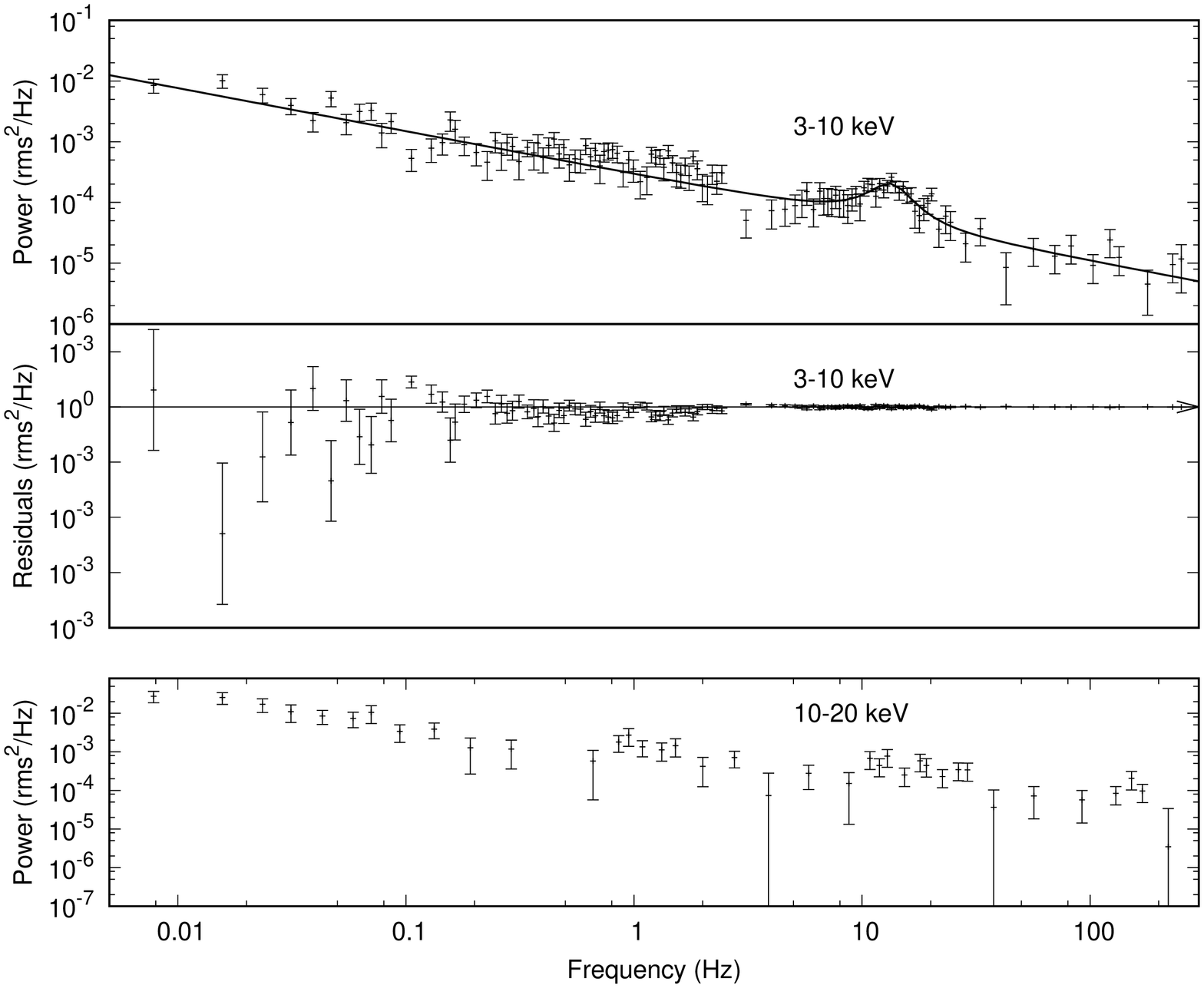} 
\vspace{1.5cm}
\caption{Top panel: Broad PN feature seen centered around 5-13.5 Hz with peak centroid frequency 13.2 $\pm$ 0.6 Hz in the energy range 3-10 keV. 
Middle panel: Corresponding residuals. Lower panel: PDS in the 10-20 keV band where no such feature was observed.}\label{fig8}
\end{figure}

Low Frequency Noise (LFN) (35-45 Hz), upper and lower kHz QPOs have been noted in the PDS of the source when it is in the 
banana state (high/soft state), while a band limited noise and also a LFN $\sim$ 9-12 Hz has been noted in the
island state which is the hard state (Olive et al. 2003). The high soft state can be modeled by
the sum of a thermal Comptonized component, a power law and a broad iron line. While reflection models have been used 
when the source is in the hard state (Di Salvo et al. 2015). 
Olive et al. (2003) interpreted spectral state transitions in the framework of a truncated accretion disk geometry
and the characteristic frequencies seen in the PDS especially the LFN $\nu$ to be related to the truncation radius of the disk. During the soft state the spectral properties (e.g. soft component
temperature and hard component temperature) and the PDS features have been found to be correlated (Olive et al. 2003), while
in the hard state they are decorrelated. In the soft state the disk is considered to be close to the NS surface with the characteristic
LFN at its maximum, i. e. disk radius at its minimum. While in the hard state the disk is truncated further away and hence the disk temperature
is low and does not contribute much to the energy spectrum. Therefore its properties or variations can be inferred from only the timing study.
Hence a comparison between the energy spectral modeling and the modeling of the power spectra can shed more information about the 
state of the source.

 Agrawal et al. (2018) has performed detailed spectral and temporal studies (HID and PDS) of the same set of observations (dataset 2)
and concluded that the source was in the banana state. For our dataset (1), the source was in the banana state
based on the HID.

\section{Results and Discussion}
\subsection{Constraining the coronal size: CCF study}
A CCF study of 4U 1705-44, between soft (3-5 keV) and hard (15-30 keV) X-ray energy bands
is reported here for the first time. Similar studies in other Z and atoll sources have been reported by
Lei et al. (2008, 2013), Sriram et al. (2019) and Malu et al. (2020). 
CCF lags obtained between soft and hard photons could be considered as the readjustment timescales of the 
soft and the hard X-ray emitting regions. Previously, our studies (Sriram et al. 2019) have shown that
the CCF lags obtained between the soft and hard energy bands could primarily be attributed to variations in the coronal structure. 
This conclusion was arrived at by initially ruling out the readjustment (viscous) timescale of the disk (assuming a Shakura-Sunyaev disk) as the
only singular causative factor as was previously considered in earlier works. The viscous timescale associated with the disk was found to be only few tens of 
seconds whereas the CCF lags obtained were of the order of few hundred seconds. Hence the CCF delay was concluded to be primarily due to the
readjustment time scale of the coronal structure with little or no contribution from the disk.
Considering the readjustment timescale of the corona to be $\beta$ times that of the disk ($\le$ 1; since corona viscosity is less than disk viscosity) 
such that the coronal readjustment velocity v$_{corona}$=$\beta$v$_{disk}$ (where v$_{disk}$ is the disk readjustment velocity), 
an equation to constrain the coronal height was derived by Sriram et al. (2019) (Equation 1).

Every light curve segment identified on the basis of the GTI were subjected to the CCF study and 
3 segments were found to display lags in dataset 1 (Table 2). All the segments that had lags were anti-correlated and two of them
exhibited hard lags and one exhibited a soft lag. Remaining segments were either highly positively correlated 
(CC $\sim$ 0.6-0.8), which was the case most of the time, or uncorrelated.
Lags obtained where 371 $\pm$ 7 s (hard lag) with CC = -0.53 $\pm$ 0.06 and
-163 $\pm$ 9 s (soft lag) with CC = -0.47 $\pm$ 0.08. For dataset 2, among the seven light curve segments that showed lags, 
three of them were hard lags while four were soft lags and these lags varied from 64 s to 573 s.

 Another aspect subject to consideration here is that for atoll sources in banana branch (high soft state) consisting of weakly magnetized
neutron stars, the disk may extend to the last stable orbit. 
Egron et al. (2013) found that the inner disk radius in the soft state of 4U 1705-44 was relatively
closer to the NS surface (10--16 Rg (Rg=GM/c$^2$)), while Lin et al. (2010) found that during the soft state, inner disk radius estimated
from the diskline model reaches 6 Rg (i. e. ISCO). Reis et al. (2009) reported that the disk of 4U 1705-44 extended towards the stellar surface with 
just a gap of 3.5 km. Since the disk is close to the last stable orbit, the observed delays are probably arising from the variability of the corona and 
such a scenario was reported by Kara et al (2019) wherein the corona structure varies with little or no movement of the disk.

Hence if the hard photons are arising from the compact corona and the
disk remains at the/close to the last stable orbit during this state then the only varying structure here would be the hard photon emitting coronal region.  
Thus using the equation below we constrain the coronal size,
\begin{equation}
H_{corona}=\Bigg[\frac{t_{lag} \dot{m}}{2 \pi R_{disk} H_{disk} \rho}-R_{disk}\Bigg] \times \beta \; cm
\end{equation}

In the equation, H$_{disk}$ = 10$^{8}$ $\alpha^{-1/10}$ $\dot{m}_{16}^{3/20} R_{10}^{9/8} f^{3/20} $ cm, 
$\rho$ = 7 $\times$ 10$^{-8}$ $\alpha^{-7/10}$ $\dot{m}^{11/20}$ $R^{-15/8}$ $f^{11/20}$  g cm$^{-3}$, f = (1-(R$_s$/R)$^{1/2}$)$^{1/4}$ 
and $\beta$ = v$_{corona}$/v$_{disk}$ (Shakura \& Sunyaev 1973, Sriram et al. 2019).

By considering the R$_{disk}$ to be at the last stable orbit ~ 12 km (following Lin et al. 2010), 
taking $\dot{m}$ $\sim$ 10$^{16}$-10$^{18}$ g s$^{-1}$ and 
$\beta$ = 0.1-0.5 ( Based on MHD simulations Manmoto et al. 1997, Pen et al. 2003, McKinney et al. 2012, e. g. similar to the
consideration made in Malu et al. 2020), we estimate the height of the coronal 
region to be 18-77 km for a lag of 371 s for $\dot{m}$ =10$^{16}$ g s$^{-1}$ and 10$^{18}$ g s$^{-1}$ respectively and $\beta$ = 0.1. 
Now by taking $\beta$ = 0.5, this H$_{corona}$ ranges from 92-384 km.

Similarly we estimate in the same way the coronal heights for causing a lag of 163 s. For 163 s lag this value is estimated to be 7-33 km ($\beta$=0.1) and 37-165 km 
($\beta$ = 0.5). For a 573 s lag as obtained from dataset 2 we estimate the coronal height to be 29-119 km for $\beta$=0.1 and 145-596 km for
$\beta$=0.5. The lowest lag of 64 s constrains the coronal height to a value of 2-12 km for $\beta$=0.1 and 11-61 km for $\beta$=0.5.

Based on our initial consideration of a physically varying coronal structure causing the lag, we can conclude that
a coronal region of few tens of km height could be condensing or expanding in these time scales.
Based on the dynamic CCF study performed in the 3-5 vs 15-30 keV energy bands we could note lags occurring within a span of less than hundred seconds
as well. These short time scale variations of few tens of seconds could mean that the disk-corona structure could be readjusting on such
short timescales as well. This leads us to consider the possibility of a model wherein such shorter timescale variations in the disk-coronal 
structure could cumulate over time and result in the overall structural variations indicated by the few hundred second delays in the CCF.

\subsection{Radius of the viscous shell and Boundary layer region}
In a previous study by Agrawal et al. (2018) a broad peaked noise component (PN) was noticed at around 1-13 Hz 
in the 3-20 keV energy band, similar to what has been noted by us in the 3-10 keV energy band. 
Agrawal et al. (2018) concluded then that since Comptonization is the dominant process in this energy band
the PN is associated with the corona. Or as suggested by Swank (2001) this feature could be produced in the disk
after which it gets modified in the corona. Another possibility could be the Boundary Layer (BL) around the NS surface
which contributes to the hot blackbody or the Comptonized component around 7-20 keV (Popham \& Sunyaev 2001, Barret et al. 2000).
Since this feature is observed only in the 3-10 keV band, we can fairly assume this to be arising from a relatively
cooler region of BL or maybe associated with the outer regions of the corona. 

Similarly Hasinger (1987) and Titarchuck et al. (2001) proposed the existence of a spherical structure/viscous shell around the NS surface
that could act as the origin for the Normal Branch Oscillations in Z sources. These QPO features then could be the oscillations within this 
shell. Now if we could extend this model by considering that the obtained broad feature centered around 5-13.5 Hz is originating from such a viscous shell, 
then by taking the centroid frequency of the PN, we can estimate the size (L$_{s}$) of this shell using the equation given by Titarchuk et al. (2001).

\begin{equation}
L_{s} = \frac{f \nu_{s}}{\nu_{ssv}}
\end{equation}

Here $\nu$$_{ssv}$ is the spherical shell viscous frequency, f is 0.5, 1/2$\pi$ for the stiff and free boundary conditions respectively
in the transition layer and $\nu$$_{s}$ is the sonic velocity given by Hasinger (1987) as,
 
\begin{equation}
\nu_{s}= 4.2 \times 10^{7} R_{6}^{-1/4} (\frac{M}{M_{\odot}} \frac{L}{L_{Edd}})^{1/8} cm \; s^{-1}
\end{equation}
where R$_6$ is the neutron star radius in units of 10$^6$ cm.

By considering a maximum luminosity of 0.5 L$_{Edd}$ (Agrawal et al. 2018) and taking the centroid frequency to be 13 Hz, we estimate the size L$_s$ to be
$\sim$ 15 km which also falls in the range of the size of corona estimated from equation 1.

If we consider a boundary layer region as proposed by Popham \& Sunyaev (2001) to be a source of hard energy photons then using the relation
proposed by them for the radius of the boundary layer, which connects the boundary layer radius and mass accretion rate, we can estimate the radius
of the source of Comptonized or hot blackbody emission.
The equation is as below,

\small
\begin{equation}
log(R_{BL} - R_{NS}) \sim 5.02 + 0.245 \Bigg[log\Big({\frac{\dot{M}} {10^{-9.85}\ \ M_{\odot} yr^{-1}}\Big)}\Bigg]^{2.19}
\end{equation}
\normalsize
Based on this equation BL radius was estimated to be 11-26 km for a $\dot{m}$ $\sim$ 10$^{16}$-10$^{18}$ g s$^{-1}$ for a NS radius R$_{NS}$ = 10 km.

\section{Conclusion}

We have performed energy dependent CCF studies of the atoll source 4U 1705-44 which revealed CCF lags of the order of few hundred seconds between the
soft and hard energy bands.
Based on the CCF and PDS study to determine the size of the corona or a sub-keplerian flow around 4U 1705-44 atoll source, 
we can conclude that a structure of few tens of km size  $\sim$ 7-77 km height ($\beta$ = 0.1) is required to explain the observed timing features. 
This is in the same range as that which was estimated for the Z source GX 17+2. 
This would hence lead to an understanding that the structural configuration in the inner region of the accretion disk and around the NS could not be drastically
different for atoll and Z sources. 
CCF studies show that there is no trend or any systematic change in 
the nature or value of the lags depending on the position on the HID. This is similar to what was seen for the Z source GX 17+2 where lags detected were random
in nature along the HB and NB (Sriram et al. 2019).

Using the  centroid frequency of the broad PN component the size of a viscous shell or a transition layer around the NS surface was estimated to be 
 $\sim$ 15 km which lies in the same range as the size estimated for the coronal structure. Non-detection of this feature in the hard band indicates that 
this feature could  be arising from a cooler region or the outer parts of the corona around the NS surface.

Understanding the long and short timescale variations in these sources could hold the key towards understanding the accretion disk corona geometry. A similar
study on various other atoll and Z sources may help in resolving the ambiguities concerning the differences between Z and atoll sources.

\section*{Acknowledgements}
We thank the Referee for providing valuable feedback that has improved the quality of the paper.
K. S. acknowledges the financial support of ISRO under AstroSat
archival Data utilization program. This publication uses data from the
AstroSat mission of the Indian Space Research Organisation (ISRO),
archived at the Indian Space Science DataCentre (ISSDC). 
M. S. acknowledges the financial support from DST-INSPIRE fellowship.
S. H. acknowledges the support from the SRF grant from CSIR-UGC.
K. S. also acknowledges the SERB Core Research Grant for the financial support.
Authors sincerely acknowledge the contribution of the LAXPC team
toward the development of the LAXPC instrument on-board the AstroSat. 
This research has made use of the data collected from the AO cycle 3 of AstroSat observations.
This work uses data from the LAXPC instruments developed at TIFR, Mumbai, and the LAXPC POC at
TIFR is thanked for verifying and releasing the data via the ISSDC
data archive. Authors thank the AstroSat Science Support Cell hosted
by IUCAA and TIFR for providing the LAXPC software that was
used for LAXPC data analysis.

\section{Declarations}

\subsection{Funding}
Not applicable .

\subsection{Conflicts of interest/Competing interests}
Not applicable.

\subsection{Availability of data and material}
Data used in this work can be accessed through the Indian Space Science Data Center (ISSDC) website 
(https://astrobrowse.issdc.gov.in/astro\_archive/archive/Home.jsp) and is also available with the authors.

\subsection{Code availability}
Not applicable.

\end{document}